\documentclass[11pt,nofootinbib,floatfix]{revtex4}
\usepackage{mathrsfs}
\usepackage{graphicx}
\usepackage{amsmath}
\usepackage{amsfonts}
\usepackage{amssymb}
\usepackage{color}

\usepackage{epsfig}
\usepackage{CJK}
\usepackage{eepic}
\usepackage{bbm}
\usepackage{dcolumn}
\usepackage{bm}
\usepackage{ulem}
\usepackage{slashbox}
\usepackage{multirow}

\newcommand{\omits}[1]{}

\def\bc{\begin{center}}

\def\ec{\end{center}}
\def\be{\begin{eqnarray}}
\def\ee{\end{eqnarray}}

\definecolor{dyellow}{rgb}{1.,0.8,.0}
\definecolor{myblue}{rgb}{.1,.1,.7}
\definecolor{dcyan}{rgb}{.0,.6,.6}
\definecolor{cyan}{rgb}{0.4,1.0,1.0}
\definecolor{dmagenta}{rgb}{0.6,0.0,0.6}
\definecolor{brown}{rgb}{0.6,0.2,0.}
\definecolor{darkblue}{rgb}{.0,.0,0.5}
\definecolor{darkred}{rgb}{0.75,0.0,0.0}
\definecolor{orange}{rgb}{1.,.6,.0}
\definecolor{dorange}{rgb}{0.8,.4,.0}
\definecolor{green}{rgb}{0.0,1.0,0.0}
\definecolor{darkgreen}{rgb}{0.0,0.6,0.0}
\definecolor{purple}{rgb}{.4,.0,.4}
\definecolor{lightgrey}{rgb}{0.7, 0.7, 0.7}
\definecolor{grey}{rgb}{0.4, 0.4, 0.4}

\def\be{\begin{equation}}
\def\ee{\end{equation}}
\def\bea{\begin{eqnarray}}
\def\eea{\end{eqnarray}}

\def\s{\sigma}

\def\l{\lambda}

\def\p{\partial}
\def\m{\mu}
\def\n{\nu}
\def\a{\alpha}

\def\>{\rangle} 
\def\<{\langle} 
\def\vev#1{\langle #1\rangle}

\def\Tr{\text{Tr}}

\begin{document}


\title{Note on R\'{e}nyi entropy of 2D perturbed free fermions}

\author{Yuan Sun$^{1}$} \email{sunyuan6@mail.sysu.edu.cn}
\author{Jia-Rui Sun$^{1}$} \email{sunjiarui@sysu.edu.cn}

\affiliation{${}^1$School of Physics and Astronomy, Sun Yat-Sen University, Guangzhou 510275, China}



\begin{abstract}
In this paper we study the R\'{e}nyi entropy of 2D massless free fermions perturbed by the $T\bar{T}$ term and the $J\bar{T}$ term at the first order perturbation. Three cases, the vacuum state of infinite size system with $T\bar{T}$ perturbation, the excited states of finite size system with $T\bar{T}$ perturbation, and the vacuum state of infinite size system with $J\bar{T}$ perturbation, are analyzed. We use the bosonization approach to calculate the perturbative expansions of R\'{e}nyi entropy. In the bosonization language the twist operator is known explicitly, with which the computation of correlators in the perturbative expansion can be simply performed. Moreover, we show that the $T\bar{T}$ and $J\bar{T}$ terms have a simple form and are similar with each other. For the first and third cases, we reproduce the known results for R\'{e}nyi entropy using the bosonization method. While for the second case, we obtain new results for the excited states.

\end{abstract}


\maketitle
\newpage
\tableofcontents
\section{Introduction}
 The quantum entanglement is a very important notion in characterizing the correlation between different subsystems in the quantum system, and it becomes a topic of great interest both in high energy physics and condensed matter physics during recent years. For a quantum system the entanglement entropy $S_A$ corresponding to a subsystem $A$ is defined via the von Neumann entropy
\be
S_A=-\Tr\rho_A\ln\rho_A,
\ee
where $\rho_A$ is   the reduced density matrix of subsystem $A$  which is defined by tracing out the degrees of freedom in the complement of $A$,  namely $\bar{A}$, as
\be
\rho_A=\Tr_{\bar{A}} \rho.
\ee
 The von Neumann entropy is not the unique way to describe the quantum entanglement, it has been generalized into the so called R\'{e}nyi entropy as
\be
S^{(n)}_A=-\frac{1}{n-1}\ln\Tr\rho^n_A
\ee
with   real parameter $n\geq 0$. The von Neumann entropy can be obtained by taking the limit $n\rightarrow 1$, namely
\be
S_A=\lim_{n\to 1}S^{(n)}_A.
\ee
Besides as a measure of entanglement, the R\'{e}nyi entropy is important in its own right. For example, if $S^{(n)}_A$ is known for all $n$, then the spectral of the reduced density matrix $\rho_A$ is  known.

Nevertheless, the entanglement entropy or R\'{e}nyi entropy is very difficult to calculate for general quantum field theory (QFT). Different methods have been developed to compute them for a variety of systems amount which free field theories or theories with additional symmetry, e.g., the conformal symmetry, are mostly concentrated. The early studies on entanglement entropy was done in the context of black hole physics with the aim to explain the possible   microscopic  origin of black hole   Bekenstein-Hawking area  entropy, where the entanglement entropy of free boson  was  calculated by discretizing the field and the area law of entanglement entropy  was  found \cite{Bombelli:1986rw,Srednicki:1993im}. Later the entanglement entropy for 2D conformal field theory (CFT)   was derived in \cite{Callan:1994py,Holzhey:1994we} by employing the powerful conformal symmetry,
and the analytic expression of entanglement entropy   was  obtained, in which an universal term was shown to  depend on the central charge $c$ of the CFT and the size of the subsystem. Meanwhile, the replica trick was also proposed to evaluate the entanglement entropy in \cite{Callan:1994py,Holzhey:1994we}. For 2D CFT another important tool, the twist operator was introduced in \cite{Calabrese:2004eu} to calculate the entanglement entropy and R\'{e}nyi entropy. The twist operator is conformal primary with conformal dimension depending on central charge $c$, this property allows people to obtain analytic results regarding the entanglement entropy and R\'{e}nyi entropy, especially in the large $c$ limit, in which  cases the CFT might have classical gravity dual. Beside utilizing the twist operator, some other methods are also developed to calculate the entanglement entropy, such as the heat kernel method \cite{Solodukhin:2011gn}, the method involving the modular Hamiltonian \cite{Casini:2010kt,Casini:2011kv} and so on.

In many cases, entanglement entropy and R\'{e}nyi entropy are considered in vacuum state. Thus it is interesting the extend the study into finite temperature or excited states, either from the CFT side \cite{Alcaraz:2011tn}, or from the holographic perspective, e.g., the first order excited states \cite{Bhattacharya:2012mi,Blanco:2013joa} and the second order excited states \cite{He:2014lfa}, or theories away from the conformal point \cite{He:2013rsa,Pang:2013lpa}. In this paper we will employ the twist operator and bosonization approaches \cite{Casini:2005rm,Casini:2007bt} to study the perturbation of R\'{e}nyi entropy for 2D free fermions perturbed by two particular  perturbations, i.e. the $T\bar{T}$ deformation, as well as the $J\bar{T}$ deformation. These kind of deformations in 2D CFT have attract much   attentions recently since they have remarkable properties from the viewpoints of both field theory and holography. Below we will review some basic aspects of the $T\bar{T}$  deformed QFT.

For any 2D Euclidean QFT with stress energy tensor $T_{\m\n}$, one can deform the theory by a composite operator $T\bar{T}(z)$ which is defined by examining a special OPE below as \cite{Zamolodchikov:2004ce}
\be\label{ttbar}
\lim_{z\to z'}(T(z)\bar{T}(z')-\Theta(z)\bar{\Theta}(z'))=\mathcal{O}_{T\bar{T}}(z')+\sum_iF_i(z-z')\p_{z'}\mathcal{O}_i(z'),
\ee
where $T=-2\pi T_{zz},\bar{T}_{\bar{z}\bar{z}}=-2\pi T_{\bar{z}\bar{z}},\Theta=2\pi T_{z\bar{z}}=2\pi T_{\bar{z}z}$, This OPE is remarkable by virtue of the fact that the second term on the right hand side of Eq.(\ref{ttbar}) only involves derivatives of operators, which leads one to define the operators $T\bar{T}(z)\equiv\mathcal{O}_{T\bar{T}}(z)$ by modulo those derivative terms.
The deformed action $S(\l)$ is parameterized by a coupling constant $\l$
\be
\int d^2zT\bar{T}(z)=\frac{\p S(\l)}{\p \l}.
\ee
Note that the stress tensor defined by using $S(\l)$ also depends on $\l$ implicitly.
Actually, for 2D integrable QFT, the $T\bar{T}$-deformation is a special case of an infinite set of integrable deformations, which means the deformed theory is still integrable in the sense that there  are still infinite number  of local
integrals of motion  \cite{Smirnov:2016lqw}.

It is known that $T\bar{T}$-deformation has good properties from the point of view of field theory. Also it will be useful in the study of the AdS/CFT correspondence \cite{Maldacena:1997re}. When generalizing the holographic duality into more general cases, it is natural to consider deformations either from the bulk side or from the boundary CFT side. Many different kinds of deformations have been investigated so far. Recently, it was suggested that the gravity with finite cutoff in the bulk is dual to the $T\bar{T}$-deformed CFT \cite{McGough:2016lol}. From renormalization group point of view, since the $T\bar{T}$-deformation is irrelevant, thus the UV properties of the field theory will be influenced, which indicates that the bulk geometry remains AdS in the interior while changes asymptotically near the boundary. The suggestion in \cite{McGough:2016lol} was  supported by the non-trivial facts that the energy spectrum of the deformed field theory are consistent with the prediction from the gravity side, and the   propagating  speed of small perturbations of the stress tensor is the same as that of metric perturbations with Dirichlet boundary conditions on the cutoff surface. Later on, further evidences from different aspects have been found in \cite{Kraus:2018xrn,Datta:2018thy,Taylor:2018xcy,Hartman:2018tkw,Wang:2018jva}.
Especially the holographic entanglement entropy \cite{Ryu:2006bv,Ryu:2006ef} was shown to match with the calculation from the dual $T\bar{T}$-deformed CFT side \cite{Donnelly:2018bef,Chen:2018eqk}. (see also \cite{Chakraborty:2018kpr,Park:2018snf}).

Motivated by the remarkable properties of $T\bar{T}$ perturbation, we will investigate the $T\bar{T}$ perturbation for free fermions by  calculating the R\'{e}nyi
entropy for vacuum state and in particular for excited states. We will employ the bosonnization approach as mentioned before. As we will  show, for free fermions this approach can also be applied to the $J\bar{T}$ perturbation which is a generalization of $T\bar{T}$ deformation. The $J\bar{T}$ deformed CFT also has an interesting holographic dual, for related discussions, see, e.g., \cite{Chakraborty:2018vja,Apolo:2018qpq,Guica:2017lia,Araujo:2018rho,Nakayama:2018ujt}.

In the remaining parts of the paper, we firstly introduce the bosonization method for calculating the R\'{e}nyi entropy of free fermions in section \ref{method}. Then in section \ref{cal} we consider the first order perturbation for the R\'{e}nyi entropy under the $T\bar{T}$ and the $J\bar{T}$ deformations, including both the vacuum state and excited states. Finally, we give a conclusion and discussions in section \ref{con}.
\section{The bosonization approach} \label{method}
In this section we  will introduce the basics of the approach to compute the R\'{e}nyi entropy by using twist operators in 2D CFT and the bosonization of free fermions \cite{Casini:2005rm,Casini:2007bt}.
Let the subsystem $A$ be a interval $(0,x_0)$ in a infinite line and the whole system staying  in the vacuum state $|0\rangle$, then the reduced density matrix  of the subsystem $A$ is
\be
\rho_A=\Tr_{\bar{A}} |0\rangle\langle 0|.
\ee
In the language of path integral, the vacuum state can be expressed as
\be
\psi[\phi_0(x)]=\int_{\phi(\tau=0,x)=\phi_0(x)}[d\phi] e^{-S_E[\phi]},
\ee
where $S_E[\phi]$ is the Euclidean action of the field $\phi(x)$, and the path integral
is performed in the lower half plane, i.e. $\tau<0,-\infty<x<\infty$. It then follows that
\be \label{zn}
\Tr\rho^n_A=\frac{Z_n}{Z_1^n} ,~~Z_n=\int[d\phi]_{R_n}e^{-S_E[\phi]},
\ee
where the integral domain is $n$-sheeted Riemann surface denoted as $R_n$, with branch cut along the  interval $A$. On the other hand, one can view the path integral in Eq.(\ref{zn}) on $R_n$ as
$n$ independent fields on a single sheet Riemann surface $R_1$
\be \label{z1}
Z_n=\int[d\phi]_{R_1}e^{-S_E[\phi_1]-...-S_E[\phi_n]}.
\ee
In addition, the $n$ fields should fulfill boundary conditions  $\phi_i(x,0^+)=\phi_{i+1}(0^-,x,),x\in[0,x_0]$. By the cyclic $Z_n$ symmetry, the boundary
conditions can be implemented by introducing the twist operators and anti-twist operators $S_n,\bar{S}_n$ sitting at two ending points $x_1=(0,0),x_2=(0,x_0)$ respectively as
\be \label{tw1}
S_n: \phi_i\rightarrow \phi_{i+1},\quad\bar{S}_n: \phi_{i+1}\rightarrow \phi_{i}.
\ee
The (anti-)twist operator is primary with conformal dimension $( \Delta_n,\bar{\Delta}_n)$
\be
\Delta_n=\bar{\Delta}_n=\frac{c}{24}\left(n-\frac{1}{n}\right).
\ee
With the twist operator inserted into Eq.(\ref{z1}), one obtains
\be
\Tr\rho^n_A=\vev{S_n(0)\bar{S}_n(x_0)}_{R_1}\sim\frac{1}{|x_0|^{4\Delta_n}}.
\ee

For general 2D CFTs, the explicit form of twist operators are unknown. However, the situation is better for free fermions in which the twist operator can be  analytically  expressed explicitly. Below we will take this advantage to study the R\'{e}nyi entropy for free fermions. To be more precisely, we consider the Dirac fermions whose action takes the form
\be \label{s0}
S_{0}=\int d^2z(\psi^*\p\psi+\bar{\psi}^*\p\bar{\psi})
\ee
with $\psi$ and $\bar{\psi}$ being the chiral and anti-chiral components of Dirac fermion respectively
\be
\Psi(z,\bar{z})=\left(\begin{array}{c}
\psi(z) \\
\bar{\psi}(\bar{z})
\end{array} \right).
\ee

In order to use the replica trick to calculate the R\'{e}nyi entropy for fermions, one firstly defines $n$ copies of fields $\psi_i,i=1,2,...,n$ on which the twist operator can act. Secondly, considering the action of twist operators on fermions as the following
\be
S_n: \psi_{i}\to\psi_{i+1}, \quad\bar{S}_n: \bar{\psi}_{i+1}\to\bar{\psi}_i,
\ee
and the condition $\psi_{n+1}=(-1)^{n-1}\psi_1$ should be satisfied, which is modified slightly from Eq.(\ref{tw1}) introduced previously. Furthermore, this action can be expressed more clearly in the matrix notation as \cite{Casini:2005rm,Casini:2007bt}
\be
S_n=\left(\begin{array}{ccccc}
0 & 1 & 0 & ... \\
\vdots & 0 & 1 & & \\
 &&...&&\\
 & & ...& 0 &1\\
(-1)^{n-1}  &  &&  & 0
\end{array} \right),
\quad\psi=\left(\begin{array}{c}
\psi_1 \\
\psi_2 \\
\vdots\\ \\
\psi_n
\end{array} \right).
\ee
This matrix can be diagonalized and the corresponding  eigenvectors are
\be \label{fo}
\tilde{\psi}_k=\frac{1}{\sqrt{n}}\sum_{j=1}^n\psi_je^{2\pi ijk/n},\quad \left(k=-
\frac{n-1}{2},-
\frac{n-1}{2}+1...,
\frac{n-1}{2}\right),
\ee
and the corresponding eigenvalues are $e^{-i2\pi k/n}$. Then the twist operator can be written as a production
\be \label{twist}
S_n=\prod_k s_k,\quad\bar{S}_n=\prod_k \bar{s}_k,
\ee
where  for  each eigenvector the action is
\be
s_k:\quad\tilde{\psi}_k\to e^{-i2\pi k/n} \tilde{\psi}_k.
\ee
The explicit form of $s_k$ can be obtained via the bosonization method, which is \cite{Datta:2014ska}
\be \label{tbo}
s_k(z,\bar{z})=:e^{\frac{ik}{n}(H_k(z)-\bar{H}_k(\bar{z}))}:,\quad\bar{s}_k(z,\bar{z})=:e^{-\frac{ik}{n}(H_k(z)-\bar{H}_k(\bar{z}))}:.
\ee
Here $H_k(z)$ and its anti-chiral counterparts are scalar fields having the following correlator
\be
 \vev{H_{k_1}(z_1)H_{k_2}(z_2)}=-\delta_{k_1k_2}\log(z_1-z_2).
\ee
Note that fields with different index $k_i$ are uncorrelated. In  addition to the twist operator, the fermion fields in diagonal representation are bosonized as
\be \label{bf1}
\tilde{\psi}_k(z)=e^{iH_k(z)},\quad\tilde{\psi}^*_k(z)=e^{-iH_k(z)},
\ee
\be
\bar{\tilde{\psi}}_k(z)=e^{-i\bar{H}_k(z)},\quad\bar{\tilde{\psi}}^*_k(z)=e^{i\bar{H}_k(z)}.
\ee
From the bosonization representation, one can easily read the conformal dimension
for each operator.
For fermion this is
\be
\Delta(\tilde{\psi}_k)=\frac{1}{2},\quad\bar{\Delta}(\tilde{\psi}_k)=0,
\ee
\be
\Delta(\bar{\tilde{\psi}}_k)=0,\quad\bar{\Delta}(\bar{\tilde{\psi}}_k)=
\frac{1}{2}.
\ee
And for twist operator
\be
\Delta(s_k)=\bar{\Delta}(s_k)=\frac{k^2}{2n^2},
\ee
which satisfy
\be
\Delta(S_n)=\sum_k\Delta(s_k).
\ee

\section{R\'{e}nyi entropy in deformed CFT}\label{cal}
In this section we will consider the perturbations of R\'{e}nyi entropy for free fermions when the $T\bar{T}$ and $T\bar{J}$ terms are presented.
\subsection{$T\bar{T}$ perturbation for vacuum state}\label{ttvac}
The action of free fermions deformed by the $T\bar{T}$ term is
\be \label{actionperf}
S_1=S_{0}+\l \int d^2z T\bar{T},
\ee
where $\l$ is a coupling constant, $T$ and $\bar{T}$ are stress tensor related to unperturbed conformal field theory $S_0$.
To calculate the R\'{e}nyi entropy, from Eq.(\ref{zn}) we have
\be
\Tr\rho^n_A=\frac{Z_n}{Z_1^n},\quad Z_n=\int_{R_n}[d\phi]e^{-S_0-\l \int_{R_n} d^2z T\bar{T}}.
\ee
 At the first order perturbation in $\l$, we have
\be \label{first}
\Tr\rho^n_A=\frac{Z_{(0)n}}{Z_{(0)1}^n}\left(1-\l\int_{R_n}d^2z\vev{T\bar{T}}_{R_n}+n\l\int_{R_1}d^2z\vev{T\bar{T}}_{R_1}\right)
\ee
with
\bea
\int_{R_n}\vev{T\bar{T}}_{R_n}&=&\frac{\int_{R_n}[d\phi](\int_{R_n}d^2z T\bar{T}(z))e^{-S_0}}{\int_{R_n}[d\phi]e
^{-S_0}}\nonumber\\
&=&\frac{\sum_{i=1}^n\vev{{(\int_{R_1} d^2z T_i\bar{T}_i(z))S_n(0)\bar{S}_n(x_0)}}_{R_1}}
{\vev{S_n(0)\bar{S}_n(x_0)}_{R_1}},
\eea
where in the last step the path integral on $R_n$ is equivalently
performed on $R_1$ via introducing $n$ copies of the original field and twist
operator as discussed in the previous section. Note that the index $i$ in $T_i\bar{T}_i$
is the replica index, and $T_i\bar{T}_i$ for each $i$ is constructed from the $i$-th copy of $n$ fields. With Eq.(\ref{first}), it is straightforward to obtain the variation of
R\'{e}nyi entropy to the first order in $\l$
\be \label{sn1}
\delta S_A^{(n)}=\frac{\l}{n-1}\left(\frac{\sum_{i=1}^n\vev{{(\int_{R_1} d^2z T_i\bar{T}_i(z))S_n(0)\bar{S}_n(x_0)}}_{R_1}}
{\vev{S_n(0)\bar{S}_n(x_0)}_{R_1}}-n\int_{R_1}d^2z\vev{T\bar{T}}_{R_1}\right).
\ee

In the next we will calculate $\delta S_A^{(n)}$ for free fermions on the plane. In this case the stress tensor for the action in Eq.$(\ref{s0})$ can be expressed as
\be
T=\psi^*\p\psi,\quad\bar{T}=\bar{\psi}^*\bar{\p}\bar{\psi}.
\ee
Note that the last term in Eq.(\ref{sn1}) vanishes since $T$ and $\bar{T}$ are
factorized on the plane, and on plane one have
\be
\vev{T}_{R_1}=\vev{\bar{T}}_{R_1}=0.
\ee
Therefore we need to calculate the 3-point correlation function from the first term of Eq.(\ref{sn1}). This was computed in \cite{Chakraborty:2018kpr} by using the Ward identity. Here we will apply the bosonization method to compute this correlator. Firstly, transforming the perturbation into the diagonal representation
by using
\bea \label{TT}
\sum_{i=1}^n T_i\bar{T}_i=\frac{1}{n} \sum_{k_1,k_2,k_3,k_4}  \tilde{\psi}^*_{k_1}\p\tilde{\psi}_{k_2}
\bar{\tilde{\psi}}^*_{k_3}\bar{\p}\bar{\tilde{\psi}}_{k_4}\delta_{k_1-k_2+k_3-k_4,nm}.
\eea
Here $m=0,\pm 1$ since $-2(n-1)\leq k_1-k_2+k_3-k_4\leq 2(n-1)$, thus $(k_1-k_2+k_3-k_4)$ can equal to $\pm n$ or zero provided $n>1$. Furthermore, only terms with $k_1=k_2$ and $k_3=k_4$ will contribute to the 3-point correlation function in Eq.(\ref{sn1}), which can be seen by considering the following correlation function (assuming $k_1\neq k_2$)
\bea \label{corr1}
\vev{\tilde{\psi}^*_{k_1}(z)\p_z\tilde{\psi}_{k_2}(z)S_n(0)\bar{S}_{n}(x_0)}.
\eea
Note that the OPE of $\tilde{\psi}^*_{k_1}$ with $\tilde{\psi}_{k_2}$ is regular since they are independent fields. Thus by substituting equations (\ref{twist}), (\ref{tbo}) and (\ref{bf1}) into Eq.(\ref{corr1}) and using the point-splitting method, one obtains
\bea
&&\lim_{z_1\to z}\p_z\vev{\tilde{\psi}^*_{k_1}(z_1)\tilde{\psi}_{k_2}(z)S_n(0)\bar{S}_{n}(x_0)}\nonumber\\
&=&\lim_{z_1\to z}\p_z\vev{e^{-iH_{k_1}(z_1)}e^{iH_{k_2}(z)}\prod_{k_3}:e^{\frac{ik_3}{n}(H_{k_3}(0)-\bar{H}_{k_3}(\bar{0}))}:\prod_{k_4}:e^{-\frac{ik_4}{n}(H_{k_4}(x_0)-\bar{H}_{k_4}(\bar{x}_0))}:}.
\eea
When $k_1\neq k_2$ the above correlator should vanish otherwise the neutrality condition for vertex operator is violated. Following the same arguements one can also find $k_3=k_4$ is necessary for $T\bar{T}$ term to make a contribution to the correlation function. Therefore one obtains
\bea \label{TT1}
\sum_{i=1}^n T_i\bar{T}_i&=&\frac{1}{n} \sum_{k_1,k_2,k_3,k_4}  \tilde{\psi}^*_{k_1}\p\tilde{\psi}_{k_2}
\bar{\tilde{\psi}}^*_{k_3}\bar{\p}\bar{\tilde{\psi}}_{k_4}\delta_{k_1-k_2+k_3-k_4,nm}
\nonumber\\&\to&\frac{1}{n} \sum_{k_1,k_2}  \tilde{\psi}^*_{k_1}\p\tilde{\psi}_{k_1}
\bar{\tilde{\psi}}^*_{k_2}\bar{\p}\bar{\tilde{\psi}}_{k_2}=\frac{1}{n}T_{tot}\bar{T}_{tot},
\eea
where the arrow "$\to$" means the relation holds in the correlator. And $T_{tot}$ (similar for $\bar{T}_{tot}$) is the
total stress tensor for $n$ copies of fields
\be
T_{tot}=\sum_{i}T_i=\sum_k \tilde{\psi}^*_{k}\p\tilde{\psi}_{k}.
\ee
This composite operator is singular and require some normal ordering prescription, we adopt the
following normal ordering
\bea \label{stress}
T(z)&=&:\psi^*(z)\p_{z}\psi(z):=\lim_{z_1\to z}\left(\frac{1}{2}(
\psi^*(z_1)\p_z\psi(z)-\p_{z_1}\psi^*(z_1)\psi(z))-\frac{1}{(z_1-z)^2}\right).
\eea
In the bosonization language, the first term in above expression is \footnote{Here the following formula for vertex operators \cite{book} is used
\be\label{nor21}
:e^{A_1}::e^{A_2}:...:e^{A_n}:=:e^{A_1+A_2+...+A_n}:=:e^{A_1+A_2+...+A_n}:\exp(\sum_{i<j}^n\vev{A_iA_j}),
\ee
where :: denote the normal ordering and $\vev{...}$ is vacuum expectation value. In the main text we suppressed the :: for vertex operator for simplicity. From the above formula it follows that on plane one have
\be \label{nor2}
:e^{ia H(z)}::e^{-ia H(w)}:=:e^{ia H(z)-ia H(w)}: e^{a^2\vev{H(z)H(w)}}=\frac{:e^{ia H(z)-ia H(w)}:}{z-w}.
\ee
 }
\bea
 \lim_{z_1\to z}\psi^*(z_1)\p_z\psi(z)&=&\lim_{z_1\to z}\p_z[e^{-i H(z_1)} e^{iH(z)}]\nonumber\\
&=&\lim_{z_1\to z}\p_z\left(\frac{e^{-i H(z_1)+iH(z)}}{z_1-z}\right)\nonumber\\
&=&\lim_{z_1\to z}\left(\frac{e^{-i H(z_1)+iH(z)}}{(z_1-z)^2}+\frac{:e^{-iH(z_1)+iH(z)}i\p_z H(z):}{z_1-z}\right)\nonumber\\
&=&\frac{1}{\epsilon^2}+\frac{1}{2}H'(z)^2-\frac{i}{2}H''(z)
\eea
with $\epsilon=z_1-z$. Similarly the second term is
\bea
&&\lim_{z_1\to z}\p_{z_1}\psi^*(z_1)\psi(z)
=-\frac{1}{\epsilon^2}-\frac{1}{2}H'(z)^2-\frac{i}{2}H''(z).
\eea
Consequently we obtain a simple form for the stress tensor
\be \label{stressb}
T_{tot}(z)=\sum_k :(\p H_k(z))^2:,
\ee
and a similar expression for anti-chiral part is
\be \label{antistressb}
\bar{T}_{tot}(\bar{z})=\sum_k :(\bar{\p} \bar{H}_k(\bar{z}))^2:.
\ee
With those results in hand, we can calculate the 3-point correlation function which contributes to the first order perturbation for R\'{e}nyi entropy in Eq.(\ref{sn1})
as follows
\bea \label{3pt1}
&&\sum_l\vev{T_l(z)\bar{T}_l(\bar{z})S_n(0)\bar{S}_n(x_0)}\nonumber\\
&=&\frac{1}{n} \sum_{k_1,k_2} \vev{  :(\p H_{k_1})^2(z)::(\bar{\p} \bar{H}_{k_2})^2(\bar{z}):\prod_{k_3}e^{\frac{ik_3}{n}(H_{k_3}(0)-\bar{H}_{k_3}(\bar{0}))}\prod_{k_4}e^{-\frac{ik_4}{n}(H_{k_4}(x_0)-\bar{H}_{k_4}(\bar{x}_0))}}.
\eea
First, considering the chiral part which is denoted as $A$ (terms involved $H(z)$, not $\bar{H}(\bar{z})$) in the correlator
\bea \label{cc}
A&\equiv&\sum_{k_1} \vev{ :(\p H_{k_1}(z))^2:
\prod_{k_2}e^{\frac{ik_2}{n}H_{k_2}(0)}\prod_{k_3}e^{-\frac{ik_3}{n}H_{k_3}(x_0)}}\nonumber\\
&=&\sum_{k_1}\vev{ :(\p H_{k_1}(z))^2:e^{\frac{ik_1}{n}H_{k_1}(0)}e^{-\frac{ik_1}{n}H_{k_1}(x_0)}} \prod_{k\neq k_1}\vev{e^{\frac{ik}{n}H_{k}(0)}e^{-\frac{ik}{n}H_{k}(x_0)}}\nonumber\\
&=&\sum_{k_1}\vev{ :(\p H_{k_1}(z))^2:e^{\frac{ik_1}{n}H_{k_1}(0)}e^{-\frac{ik_1}{n}H_{k_1}(x_0)}}  \prod_{k\neq k_1}\left(-\frac{1}{x_0}\right)^{\frac{k^2}{n^2}}\nonumber\\
&=&\sum_{k_1}\vev{ :(\p H_{k_1}(z))^2:e^{\frac{ik_1}{n}H_{k_1}(0) -\frac{ik_1}{n}H_{k_1}(x_0)}}  \prod_{k}\left(-\frac{1}{x_0}\right)^{\frac{k^2}{n^2}} ,
\eea
where Eq.(\ref{nor21}) is used in the last step. Using Wick theorem, the correlator in last line can be calculated as
\bea
&& \p H_{k_1}(z):e^{\frac{ik_1}{n}H_{k_1}(0) -\frac{ik_1}{n}H_{k_1}(x)}: \nonumber\\
&=&\frac{ik_1}{n}\p_z[\vev{H_{k_1}(z)(H_{k_1}(0)-H_{k_1}(x_0))}]:e^{\frac{ik_1}{n}H_{k_1}(0) -\frac{ik_1}{n}H_{k_1}(x_0)}:+\text{normal ordering}\nonumber\\
&=&\frac{ik_1}{n}\frac{x_0}{z(z-x_0)}:e^{\frac{ik_1}{n}H_{k_1}(0) -\frac{ik_1}{n}H_{k_1}(x_0)}:
+\text{normal ordering}.
\eea
By iterating this
procedure twice, it then follows that
\bea  \label{nor3}
&& :(\p H_{k_1}(z))^2::e^{\frac{ik_1}{n}H_{k_1}(0) -\frac{ik_1}{n}H_{k_1}(x_0)}: \nonumber\\
&=&-\frac{k_1^2}{n^2}\frac{x_0^2}{z^2(z-x_0)^2}:e^{\frac{ik_1}{n}H_{k_1}(0) -\frac{ik_1}{n}H_{k_1}(x_0)}:
+\text{normal ordering}.
\eea
Finally  substituting  Eq.(\ref{nor3}) into Eq.(\ref{cc}), yields
\be
A=\prod_{k}\left(-\frac{1}{x_0}\right)^{\frac{k^2}{n^2}}
 \sum_{k_1}\left(-\frac{k_1^2}{n^2}\right)\frac{x_0^2}{z^2(z-x_0)^2}=-2\Delta_n
 \frac{x_0^2}{z^2(z-x_0)^2}\left(-\frac{1}{x_0}\right)^{2\Delta_n}.
\ee
Following the same  steps as derived above for chiral part of the correlator,
one can obtain the result for the anti-chiral part denoted as $\bar{A}$, which turns out to be
\bea
\bar{A} \equiv  \sum_{k_1}\vev{:(\bar{\p}\bar{H}(\bar{z}))^2:\prod_{k_2}
e^{-\frac{ik_2}{n}\bar{H}_{k_2}(\bar{0})}
\prod_{k_3}
e^{\frac{ik_3}{n}\bar{H}_{k_3}(\bar{x}_0)}}=-2\Delta_n\frac{\bar{x}_0^2}{(\bar{z}-\bar{x}_0)^2\bar{z}^2}
\left(-\frac{1}{\bar{x}_0}\right)^{2\Delta_n}.
\eea
Combining the chiral and anti-chiral  parts, we obtain the final result
\bea \label{3pt1}
&&\sum_l\vev{T_l(z)\bar{T}_l(\bar{z})S_n(0)\bar{S}_n(x_0)}=\frac{1}{n}
\frac{4\Delta^2_n}{|z-x_0|^4|z|^4|x_0|^{4\Delta_n-4}},
\eea
which has the correct form for 3-point correlation function in CFT. To obtain the first order perturbation of R\'{e}nyi entropy, one can substitute the above correlator into the integral in Eq.(\ref{sn1}). This integral is divergent and a UV cutoff is needed to deal with the divergence, for detailed discussions regarding this integral, refer to \cite{Chakraborty:2018kpr}.

Until now we have discussed the case where the system is of infinite range, and stays in its vacuum state. Along the same lines, one can consider the system with finite size or finite temperature, in which the correlators on cylinder are involved. we will not discuss these two cases explicitly in the present paper. Instead, we would like to consider a related situation involving excited states with finite size by applying the bosonization approach.
\subsection{$T\bar{T}$ perturbation for excited state}
In this subsection we will analyze the R\'{e}nyi entropy of exited states for a system with finite size. For finite size and zero temperature. The R\'{e}nyi entropy of general CFT with $T\bar{T}$ deformation has been considered in \cite{Chen:2018eqk}. The case we considered here is concentrated on free fermions as  previous subsection, we will show that bosonization method is still applicable.

Following \cite{Sarosi:2016oks} the excited states on cylinder can be prepared as
\be
|\Upsilon\rangle=\lim_{\xi,\bar{\xi}\to -\infty}\Upsilon(\xi,\bar{\xi})|0\rangle,
\ee
where $\xi= t+i\s,(\s\sim\s+2\pi,-\infty<t<\infty)$ is coordinate on the infinite cylinder, and $\Upsilon(\xi,\bar{\xi})$ is a local operator inserted at infinite past to create an incoming excited state. Let the subsystem $A$ be the region $\a<\s<\beta ,t=0$, then the corresponding reduced density matrix is
\be
\rho_A=\Tr_{\bar{A}}|\Upsilon\rangle\langle\Upsilon|.
\ee
As before, to compute the R\'{e}nyi entropy, firstly, we have
\be
\Tr\rho^n_A=\lim_{\xi,\bar{\xi}\to -\infty}\frac{\int_{C_n} D\phi\prod_{i=1}^n\Upsilon_i^\dagger(-\xi)\Upsilon_i(\xi)e^{-S_1}}{[\int_{C_1} D\phi \Upsilon^\dagger(-\xi)\Upsilon(\xi) e^{-S_1}]^n},
\ee
where $S_1$ is the perturbed action (\ref{actionperf}), and $C_n$ is $n$ copies of cylinder pasted along the interval $A$ in a cyclic order. Then keeping the first order of $\l$ in the variance  of the R\'{e}nyi entropy between the perturbed and unperturbed excited states
\bea \label{ds2}
\delta S_A^{(n)}&=&\frac{\l}{n-1} \lim_{\xi,\bar{\xi}\to- \infty}\left(\l\frac{\int_{C_n}d^2z\vev{T\bar{T}(z)\prod_{i=1}^n\Upsilon_i^\dagger(-\xi)\Upsilon_i(\xi)}_{C_n}}{\vev{\prod_{i=1}^n\Upsilon_i^\dagger(-\xi)\Upsilon_i(\xi)}_{C_n}}\right. \nonumber\\
&&\left.-n\l\frac{ \int_{C_1}d^2z\vev{T\bar{T}(z)\Upsilon^\dagger(-\xi)\Upsilon(\xi)}_{C_1}}{\vev{ \Upsilon^\dagger(-\xi)\Upsilon(\xi)}_{C_1}}\right),
\eea
which reduces to Eq.(\ref{sn1}) when $\Upsilon$ is the identity operator.

In the following we will choose $\Upsilon(z)=\psi(z)$, in which case the bosonization method can be used and we can directly bosonize the field  $\psi(z)$. The vacuum state may also be excited by other operators constructed from the fermion fields $\psi(z)$,  for example the operator $\bar{\Psi}\Psi=\bar{\psi}^*\psi+\psi^*\bar{\psi}$  as discussed in \cite{Nozaki:2015mca,Caputa:2015qbk}. In addition, for free fermions, $\delta S_A^{(n)}$  with all integer $n$ can in principle be calculated, for simplicity  we only consider the second  R\'{e}nyi entropy ($n=2$) and will show that the bosonization formula is still applicable and can simplify the calculation.

To compute $\delta S_A^{(n=2)}$, let us map the $n$ copies of cylinder $C_n$ to   $n$ copies of planes denoted as $R_n$ via the exponential map \cite{Sarosi:2016oks}
\be \label{expo}
z=e^\xi.
\ee
Here $R_n$ is pasted along the interval on unit circle with ending points $u=e^{i\a},v=e^{i\beta}$ on each plane, which is different from the situations in previous section. Under the exponential map, the stress tensor of the field is transformed via \footnote{The Hermitian conjugate have a non-trivial action on coordinates on cylinder with Euclidean time $t$ as $
t\to-t,~\s\to \s
$,
By exponential map this leads to  coordinates on plane transform under Hermitian conjugate as $z\to\frac{1}{z^*}$. Therefore
\be
[\psi(\xi)]^\dagger=[z^{\frac{1}{2}}\psi^{(p)}(z)]^\dagger=\left(\frac{1}{z^*}\right)^{\frac{3}{2}}\psi^{(p)*}\left(\frac{1}{z^*}\right),
\ee
where in the last step we have used the definition
\be \label{hcd}
 [\psi^{(p)}(z)]^\dagger=\left(\frac{1}{z^*}\right)\psi^{(p)*}\left(\frac{1}{z^*}\right),
\ee
}
\be
T(\xi)=z^2T^{(p)}(z)-\frac{c}{24},~~\psi(\xi)=\left(\frac{\p z}{\p \xi}\right)^
{\frac{1}{2}}
\psi^{(p)}(z)=z^{\frac{1}{2}}\psi^{(p)}(z),
\ee
where $c=1$ is the central charge of underlying Dirac fermion and index $(p)$ denotes the quantities defined on plane. The anti-chiral counterpart have similar expressions. As for the R\'{e}nyi entropy, notice that the nominator in the first term of
Eq.(\ref{ds2}) is the most complicate correlator in $\delta S_A^{(n)}$ which can be rewritten as
\bea \label{firstt}
&&\lim_{\xi,\bar{\xi}\to -\infty}\int_{C_n}d^2\zeta\vev{T\bar{T}(\zeta)\prod_{i=1}^n\psi_i^\dagger(-\xi)\psi_i(\xi)}_{C_n}\nonumber\\
&=& \lim_{\substack{w_1\to0\\w_2\to \infty}}w_2^n\int_{R_n}\frac{d^2z}{z\bar{z}}\vev{(z^2T^{(p)}(z)-\frac{c}{24})(\bar{z}^2\bar{T}^{(p)}(\bar{z})-\frac{c}{24})\prod_{i=1}^n\psi_i^{(p)*}(w_2)\psi^{(p)}_i(w_1)}_{R_n},
\eea
where $w_1=1/w_2$ and the prefactor $w_2^n$ comes from the definition Eq.(\ref{hcd}). Expanding the integrand and focusing first on the correlator involving both $T^{(p)}$ and $\bar{T}^{(p)}$, while the remaining correlators in the integrand can be evaluated in a similar manner
\bea \label{r1c}
&& \lim_{\substack{w_1\to0\\w_2\to \infty}}w^n_2 \int_{R_n}d^2z|z|^2\vev{(T^{(p)}\bar{T}^{(p)}(z)\prod_{i=1}^n\psi_i^{(p)*}(w_{1})\psi_i^{(p)}(w_{2})}_{R_n}\nonumber\\
&=&
\lim_{\substack{w_1\to0\\w_2\to\infty }}w^n_2 \int_{R_1}d^2z|z|^2\sum_{i=1}^n\vev{(T_i^{(p)}\bar{T}_i^{(p)}(z)\prod_{i=1}^n\psi_i^{(p)*}(w_1)\psi^{(p)}_i(w_2)S_n(u)\bar{S}_n(v)}_{R_1},
\eea
where the correlator evaluated on $n$-sheet plane $R_n$ is written equivalently on a single plane $R_1$ with
twist operator inserted. Note that the index $i$ in the first line represents the $i$-th sheet where the operator is inserted in, while the index $i$ in the second line means the $i$-th copy of $n$ fields (We will omit the index $(p)$ hereafter since only quantities on plane are needed in what follows). Now the correlator in the second line of above equation is similar with the one in Eq.(\ref{3pt1}) and can also be evaluated by using bosonization method. After transforming the fermion field into diagonal representation by employing Eq.(\ref{fo}), we obtain (for $n=2$),
\be
\psi_1(z)=-\frac{i}{\sqrt{2}}(\tilde{\psi}_+(z)-\tilde{\psi}_-(z)),~~
\psi_2(z)=-\frac{1}{\sqrt{2}}(\tilde{\psi}_+(z)+\tilde{\psi}_-(z))
\ee
with the abbreviated notation notation $\tilde{\psi}_\pm\equiv\tilde{\psi}_{\pm 1/2}$. Subsequently the product of fermion fields in the last line of Eq.(\ref{r1c}) becomes
\bea
\prod_{i=1}^2 \psi_i(z) =\frac{i}{2}(\tilde{\psi}^2_+(z)-
\tilde{\psi}^2_-(z)+2\tilde{\psi}_+(z)\tilde{\psi}_-(z))=i\tilde{\psi}_+(z)\tilde{\psi}_-(z),
\eea
where in the first step the anti-commutative property for fermionic fields is utilized, and in the second step the expression is simplified because of the terms  $\tilde{\psi}^2_-(z)$ and $\tilde{\psi}^2_+(z)$  vanish, which can be seen from the vanishing OPE as $z$ approaching $w$
\be
e^{iH_\pm(z)}e^{iH_\pm(w)}\sim e^{i(H_\pm(z)+H_\pm(w))}(z-w),~~\tilde{\psi}_\pm(z)=e^{iH_\pm(z)}.
\ee
Thus in terms of boson fields $H_\pm(z)$ we obtain
\be
\prod_{i=1}^2 \psi_i(w_1)=ie^{i H_+(w_1)}e^{iH_-(w_1)}
\ee
and for Hermitian conjugate fields
\be \label{w22}
 \prod^2_{i=1} \psi^{*}_i(w_2)=-i e^{-i H_+(w_2)}e^{-iH_-(w_2)}.
\ee

In the present case the $T\bar{T}$ term in the last line of Eq.(\ref{r1c}) still satisfies Eq.(\ref{TT1}) for the same reason as has been discussed there. Thus the calculation can be simplified by applying the bosonized form of stress tensor in Eq.(\ref{stressb}). Combining the above expressions with twist operators in Eq.(\ref{tbo}) we can now compute the correlator in Eq.(\ref{r1c}) as
\bea\label{extt}
\lim_{\substack{w_1\to0 \\w_2\to \infty}}w^2_2\sum_{j=1}^{n=2}\vev{(T_j\bar{T}_j(z)\prod_{i=1}^{n=2}\psi_i^{*}(w_2)\psi_i(w_1)S_n(u)\bar{S}_n(v)}_{R_1}\equiv\lim_{\substack{w_1\to0 \\w_2\to \infty}}w^2_2 B\bar{B},
\eea
where the chiral and anti-chiral parts have been denoted as $B$ and $\bar{B}$ respectively. The anti-chiral part $\bar{B}$ has the same form as that of Eq.(\ref{3pt1}), since the excitation of chiral fields $\psi(z)$ will only affect the chiral part of the correlator while keeping the anti-chiral part
unchanged. The explicit expression of $\bar{B}$ is
\bea\label{bbar}
\bar{B}=-\frac{1}{8}\left(\frac{\bar{u}-\bar{v}}{(\bar{z}-\bar{u})(\bar{z}-\bar{v})}\right)^2\left(\frac{1}{(\bar{u}-\bar{v})}\right)^{\frac{1}{8}},
\eea
And the chiral-part $B$  can be written as
\bea
B\equiv\sum_{k} \langle(:\p H_{k}(z))^2:e^{-i H_+(w_2)}e^{-iH_-(w_2)}e^{i H_+(w_1)}e^{iH_-(w_1)}
\prod_{k_1} e^{\frac{ik_1}{n}H_{k_1}(u)} \prod_{k_1} e^{-\frac{ik_2}{n}H_{k_2}(v)} \rangle, \nonumber\\
\eea
 which can be further decomposed into the sum of the following two terms
\bea
B_1\equiv\vev{(\p H_+(z))^2e^{-iH_+(w_2)} e^{iH_+(w_1)} e^{\frac{i}{4}H_+(u)} e^{-\frac{i}{4}H_+(v)}
 }
  \vev{ e^{-iH_-(w_2)}e^{iH_-(w_1)} e^{-\frac{i}{4}H_-(u)} e^{\frac{i}{4}H_-(v)}  } ,
\eea
\bea
B_2\equiv\vev{e^{-iH_+(w_2)} e^{iH_+(w_1)}e^{\frac{i}{4}H_+(u)} e^{-\frac{i}{4}H_+(v)}   }
\vev{ (\p H_-(\bar{z}))^2e^{-iH_-(w_2)}  e^{iH_-(w_1)} e^{-\frac{i}{4}H_-(u)}e^{\frac{i}{4}H_-(v)}  } .
\eea
By using a similar normal ordering as in Eq.(\ref{nor3}), a straightforward calculation yields
\be
B_1=-\left(\frac{w_1-w_2}{(z-w_1)(z-w_2)}+\frac{u-v}{4(z-u)(z-v)}\right)^2\left(\frac{1}{w_2-w_1}\right)^2\left(\frac{1}{u-v}\right)^{\frac{1}{8}},
\ee
\be
B_2=-\left(\frac{w_1-w_2}{(z-w_1)(z-w_2)}-\frac{u-v}{4(z-u)(z-v)}\right)^2\left(\frac{1}{w_2-w_1}\right)^2\left(\frac{1}{u-v}\right)^{\frac{1}{8}}.
\ee
Consequently the chiral-part is
\be
B=B_1+B_2=-2\left(\frac{1}{u-v}\right)^{\frac{1}{8}}\left(\frac{1}{w_2-w_1}\right)^2\left[\left(\frac{w_1-w_2}{(z-w_1)(z-w_2)}\right)^2+\left(\frac{u-v}{4((z-u)(z-v)}\right)^2\right].
\ee
Taking the limit $w_1\to 0,w_2\to\infty$, we have
\be
\lim_{\substack{w_1\to\infty\\w_2\to 0}}w_2^2B=-2\left(\frac{1}{u-v}\right)^{\frac{1}{8}
}\left[\frac{1}{z^2}+\left(\frac{u-v}{4((z-u)(z-v)}\right)^2\right],
\ee
where the second term is just the Hermitian conjugate of $\bar{B}$ in
(\ref{bbar}), and is presented no matter the whole state is vacuum or excited, while the first term appears due to the excitation. So far we have obtained the correlator in Eq.(\ref{r1c}) which belongs to one of four correlators in Eq.(\ref{firstt}). The rest three correlators are can be  calculated following the same procedures, we just list their results as follows (for $n=2$)
\bea
\lim_{\substack{w_1\to0 \\w_2\to \infty}}w^n_2\sum_{j=1}^n\vev{T_j\prod_{i=1}^n\psi_i^{*}(w_1)\psi_i(w_2)S_n(u)\bar{S}_n(v)}_{R_1}=\lim_{\substack{w_1\to0 \\w_2\to \infty}}\frac{w_2^2B}{(\bar{u}-\bar{v})^{\frac{1}{8}}},
\eea
\bea
\lim_{\substack{w_1\to0 \\w_2\to \infty}}w^n_2\sum_{j=1}^n\vev{\bar{T}_j\prod_{i=1}^n\psi_i^{*}(w_1)\psi_i(w_2)S_n(u)\bar{S}_n(v)}_{R_1}= \frac{\bar{B}}{(u-v)^{\frac{1}{8}}},
\eea
\bea \label{ppss}
\lim_{\substack{w_1\to\infty\\w_2\to 0}} w^n_2\vev{ \prod_{i=1}^n\psi_i^{*}(w_1)\psi_i(w_2)S_n(u)\bar{S}_n(v)}_{R_1}= \frac{1}{|u-v|^{\frac{1}{4}}}.
\eea

Next, let us considering the denominator of the first term of $\delta S_A^{(n)}$ in Eq.(\ref{ds2}). By doing the similar calculations as above, the results for $n=2$ turn out to be
\bea
&& \lim_{\xi,\bar{\xi}\to- \infty}
 \vev{\prod_{i=1}^n\Upsilon_i^\dagger(-\xi)\Upsilon_i(\xi)}_{C_n} =
\lim_{\substack{w_1\to0\\w_2\to\infty }}w^n_2   \vev{ \prod_{i=1}^n\psi_i^{*}(w_1)\psi^{}_i(w_2)S_n(u)\bar{S}_n(v)}_{R_1}=\frac{1}{|u-v|^{\frac{1}{4}}}.
\eea
which is just equal to Eq.(\ref{ppss}).
The remaining correlators in $\delta S_A^{(n)}$, i.e. the second term in Eq.(\ref{ds2}), are evaluated on a single cylinder $C_1$, thus no twist operators are needed. To compute these correlators, we first map the cylinder $C_1$ to plane $R_1$ by the exponential map Eq.(\ref{expo}), then we directly make bosonization for these fields without performing the Fourier transformation, since there is only one copy of the field. After performing these procedures, we find the denominator is equal to one, and the nominator is
\bea
&&\lim_{\xi,\bar{\xi}\to -\infty}\int_{C_1}d^2\zeta\vev{T\bar{T}(\zeta)\psi^\dagger(-\xi)\psi(\xi)}_{C_1}\nonumber\\
&=& \lim_{\substack{w_1\to0\\w_2\to \infty}}w_2\int_{R_1}\frac{d^2z}{z\bar{z}}\vev{(z^2T(z)-\frac{c}{24})(\bar{z}^2\bar{T}(\bar{z})-\frac{c}{24})\psi^{*}(w_2)\psi(w_1)}_{R_1}
\eea
with the correlator computed by the bosonization method
\be
\lim_{\substack{w_1\to0\\w_2\to \infty}}w_2  \vev{(z^2T(z)-\frac{c}{24})(\bar{z}^2\bar{T}(\bar{z})-\frac{c}{24})\psi^{*}(w_2)\psi(w_1)}_{R_1}=\frac{cz^2}{24}+\left(\frac{c}{24}\right)^2
\ee
So far we have obtained  all of the correlators in the right hand side of Eq.(\ref{ds2}). However the integrals in Eq.(\ref{ds2}) have not been calculated explicitly. The  integrand is divergent at some points thus some regularization might be needed, and we will leave these problems for future investigation.
\subsection{$J\bar{T}$ perturbation}
In this subsection we will consider another kind of perturbation which is called the $J\bar{T}$ perturbation. We take the same setup for the present case as in Section~\ref{ttvac}, i.e. the whole system is in its vacuum state living on an infinite line,  and the subsystem is a finite interval.

As described in \cite{Guica:2017lia} for free fermions the $J\bar{T}$ perturbation can be constructed as follows. Let us consider the action Eq.(\ref{s0}) which possesses internal symmetries as well as the conformal symmetry. One of the internal symmetry is the $U(1)$ symmetry
of $\psi$: $\psi\to e^{ia}\psi,\psi^*\to e^{-ia}\psi^*$, while keep the $\bar{\psi}$ unchanged. The current associated with this symmetry is
\be
J=J_z=\psi^*\psi.
\ee
Combining this current with the anti-chiral part of the stress tensor
\be
\bar{T}=\bar{\psi}^*\bar{\p}\bar{\psi},
\ee
one  obtains the $J\bar{T}$ term for the unperturbed action $S_0$. Then the perturbed action can be expressed as
\be
S=S_0+\l\int d^2z J\bar{T}.
\ee
In order to compute the first order perturbation of the R\'{e}nyi entropy, one just need to replace the $T\bar{T}$ term in Eq.(\ref{sn1}) with the $J\bar{T}$ term, thus the resulting correlator to be calculated is
\be \label{JTc}
\sum_l\vev{J_l(z)\bar{T}_l(\bar{z})S_n(0)\bar{S}_n(x_0)}.
\ee
To proceed, one can also employing the bosonization method to evaluate this correlator.
As we have done  before we first transform the perturbation term to diagonal representation
\bea \label{JT1}
\sum_{l=1}^n J_l\bar{T}_l(z)
&=&\frac{1}{n} \sum_{k_1,k_2,k_3,k_4}  \tilde{\psi}^*_{k_1} \tilde{\psi}_{k_2}
\bar{\tilde{\psi}}^*_{k_3}\bar{\p}\bar{\tilde{\psi}}_{k_4}\delta_{k_1-k_2+k_3-k_4,mn},~~m=\pm1,0\nonumber\\
&\to&\frac{1}{n}  \sum_{k_1,k_3}  \tilde{\psi}^*_{k_1} \tilde{\psi}_{k_1}
\bar{\tilde{\psi}}^*_{k_3}\bar{\p}\bar{\tilde{\psi}}_{k_3} \equiv\frac{1}{n}J_{tot}
\bar{T}_{tot},
\eea
which is similar with Eq.(\ref{TT1}). The arrow $\to$ means that the  replacement holds in the correlator. Here we  have introduced the notation
\be
J_{tot}=\sum_l J_l=\sum_k \tilde{\psi}^*_k \tilde{\psi}_k,
\ee
which can be bosonized as
\bea
\tilde{\psi}^*_k \tilde{\psi}_k(z)=\lim_{\epsilon\to0}\left(e^{-i H_k(z+\epsilon)}e^{iH_k(z)}-\frac{1}{\epsilon}\right)=\lim_{\epsilon\to0}\frac{ e^{-i H_k(z+\epsilon)+iH_k(z)}-1}{\epsilon}=-i\p H_k(z)
\eea
Therefore in the language of bosonization, the perturbation term takes the form as
\be
\sum_{l=1}^n J_l\bar{T}_l(z)=-\frac{i}{n}\sum_{k_1,k_2}\p H_{k_1}(z):(\bar{\p} \bar{H}_{k_2})^2(\bar{z}):.
\ee
which have the similar form as the case of $T\bar{T}$ perturbation.

 Considering  the first order perturbation
\bea \label{3pt2}
&&\sum_l\vev{J_l(z)\bar{T}_l(\bar{z})S_n(0)\bar{S}_n(x_0)}\\
&=&\frac{-i}{n} \sum_{k_1,k_2} \vev{  \p H_{k_1}(z):(\bar{\p} \bar{H}_{k_2})^2(\bar{z}):\prod_{k_3}e^{\frac{ik_3}{n}(H_{k_3}(0)-\bar{H}_{k_3}(\bar{0}))}\prod_{k_4}e^{-\frac{ik_4}{n}(H_{k_4}(x_0)-\bar{H}_{k_4}(\bar{x_0}))}},\nonumber
\eea
which can be factorized into the  chiral part and the anti-chiral part. The anti-chiral part is the same as the $T\bar{T}$ case as discussed in section \ref{ttvac}, and the chiral part turn out to be zero which can be seen as follows
\bea
&&\sum_{k_1} \vev{ \p H_{k_1}(z)
\prod_{k_2}e^{\frac{ik_2}{n}H_{k_2}(0)}\prod_{k_3}e^{-\frac{ik_3}{n}H_{k_3}(x_0)}}\nonumber\\
&=&\sum_{k_1}\vev{  \p H_{k_1}(z)e^{\frac{ik_1}{n}H_{k_1}(0)}e^{-\frac{ik_1}{n}H_{k_1}(x_0)}} \prod_{k\neq k_1}\vev{e^{\frac{ik}{n}H_{k}(x)}
e^{-\frac{ik}{n}H_{k}(0)}}\nonumber\\
&=&-\sum_{k_1}\frac{ik_1}{n}\frac{x_0}{z(z-x_0)}   \prod_{k}\left(-\frac{1}{x_0}\right)^{\frac{k^2}{n^2}} =0,
\eea
where in the last step the summation over $k_1$ cancels to zero. This can be understood as follows since the the $J$ is a primary operator, and the first order contribution to R\'{e}nyi entropy vanishes if the perturbation operator is primary. This is in contrast with $T\bar{T}$ perturbation where $T$ is not primary and thus the first order contribution to R\'{e}nyi entropy is nonzero. Therefore to see the nonvanishing contribution to R\'{e}nyi entropy for $J\bar{T}$ perturbation one should go to higher order perturbation.
\section{Conclusion}\label{con}
In this paper we investigated the $T\bar{T}$ as well as the $J\bar{T}$ perturbation of 2D massless Dirac fermions which is a free CFT. We mainly studied  the perturbation of R\'{e}nyi entropy by using the bosonization approach for three cases. In the first case, we considered the R\'{e}nyi entropy to the first order of a vacuum state on plane, and showed  that the $T\bar{T}$ term can be written in a simple form in terms of the bosonic fields as presented in Eq.(\ref{stressb}) and Eq.(\ref{antistressb}). By using this bosonic form of the stress tensor, we reproduced the 3-point correlator appeared in the first order variation of the R\'{e}nyi entropy, which has been previously computed via Ward identity in \cite{Chakraborty:2018kpr}. The advantages of the bosonization method is that it allows us to handle more complicated correlators in a simple way, while the Ward identity is insufficient to obtain the correlators, for example, the second case we considered. In this case, we generalized the above discussion for vacuum state to excited states on cylinder, where the states are excited by the chiral operator $\psi(z)$. In this case we explicitly analyzed the perturbation of second R\'{e}nyi entropy $\delta S_A^{(n=2)}$  at the first order perturbation. Although for excited states, the correlators are more complicated than the ones as for the vacuum state, however, we showed that the bosonization method is still useful, and the corresponding correlators can be simply evaluated using this method. In the last case, we considered the $J\bar{T}$ perturbation for vacuum state on the plane. The bosonized form of $J\bar{T}$ term presented in Eq.(\ref{JT1}) exhibits a similar form with the $T\bar{T}$ term. Then applying the bosonization approach, we showed that the R\'{e}nyi entropy vanishes at the first order perturbation, which is consistent with the fact that for vacuum state the first order  perturbation of single interval R\'{e}nyi entropy caused by primary operators always vanish.

In the discussion for the R\'{e}nyi entropy, we only analyzed the cases for the single interval subsystem, it is also tractable to generalize the situation into multi-interval cases, e.g., studying the perturbation of mutual information for free fermions which involves two-interval subsystem \cite{Morrison:2012iz}. In addition, it is also interesting to consider finite temperature and finite size system. This system for free fermion without perturbation has been discussed in \cite{Ogawa:2011bz,Azeyanagi:2007bj}.

\section*{Acknowledgement}
We would like to thank Soumangsu Chakraborty for helpful discussions. This work was supported by the National Natural Science Foundation of China (No.~11675272), the Open Project Program of State Key Laboratory of Theoretical Physics, Institute of Theoretical Physics, Chinese Academy of Sciences, China (No.~Y5KF161CJ1) and the Fundamental Research Funds for the Central Universities.





\end{document}